\documentclass[12pt]{article}
\usepackage{amssymb,amsmath,cite,epsfig}
\usepackage[top=2.5cm, bottom=2.5cm, left=2.5cm, right=2.5cm]{geometry}
\usepackage{indentfirst}
\usepackage{subfig}
\begin{document}
\begin{titlepage}
\begin{flushright}
\end{flushright}
\vspace*{3cm}
\begin{center}
{\Large \textsf{\textbf{Thermodynamic Properties of Yukawa-Schwarzschild Black Hole in Non-commutative Gauge Gravity}}}
\end{center}
\par \vskip 5mm
\begin{center}
{\large \textsf{Slimane Zaim and Hadjar Rezki}}\\\vskip
5mm
D\'{e}partement de Physique, Facult\'{e} des Sciences de la Mati\`{e}re,\\
Universit\'{e} de Batna $1$, Algeria. \\\vskip
5mm
Email: zaim$69$slimane$@$yahoo.com
\end{center}
\par \vskip 2mm
\begin{center} {\large \textsf{\textbf{Abstract}}}\end{center}
\begin{quote}
\setlength{\parindent}{3ex}{
In this paper we construct a non-commutative gauge theory for the deformed metric corresponding to the modified structure of a gravitational field in the case of Yukawa-Schwarzschild non-commutative space-time. The thermodynamic properties and corrections to the gravitational force on the horizon of a non-commutative Yukawa-Schwarzschild black hole are analysed.}
\end{quote} \vspace*{2cm}
\begin{quote}
\textbf{\sc Pacs numbers}: 11.10.Nx, 11.15.-q, 03.65.Pm
\end{quote} \vspace*{2cm}
\end{titlepage}

\section{Introduction}

\setlength{\parindent}{3ex}
General relativity gives rise to the concept of black holes as a real physical phenomenon. Interest in this phenomenon has dramatically increased since Hawking discovered a mechanism by which black holes can radiate \cite{1}. This mechanism involves quantum-mechanics processes near the black hole horizon. In fact Hawking showed that a black hole radiates a spectrum of particles which is quite analogous to thermal black body radiation. Thus Hawking radiation emerges as a nontrivial consequence of combining gravity and quantum mechanics. Mathematically this phenomenon was treated in Refs. \cite{1,2}. For a given static and spherically-symmetric metric, the black hole
temperature ($T_{H}$) is proportional to the surface gravity ($k$) according to the relation $T_{H}=\frac{k}{2\pi}$. This is the interpretation of Hawking radiation. This formalism can also be applied to a static and spherically symmetric non-commutative Yukawa-Schwarzschild space-time.

Quantum mechanics is based on the notion of commutation relations between position and momentum. We can think of a different approach in which one enforces commutation relations between position coordinates themselves as
well as momentum coordinates. In analogy to the Heisenberg uncertainty relations between position and momentum, a new set of uncertainty relations appears between position coordinates themselves as well as momentum
coordinates. This idea results in the concept of quantum gravity since quantifying space-time leads to quantifying gravity. Non-commutativity is mainly motivated by string theory, being a limit in the presence of a
background field \cite{1, 2, 3}. Here one uses a gauge field theory with star products and Seiberg-Witten maps. The non-commutative space-time is characterized by the coordinate operators $\hat{x}^\mu$ satisfying the
following communication relation:
\begin{equation}\label{eq:1}
\left[\hat{x}^\mu,\hat{x}^\nu\right]_\star=i\theta^{\mu\nu}\,,
\end{equation}
where $\theta^{\mu\nu}$ is an anti-symmetric real matrix which determines the fundamental cell discretization of space-time much in the same way as the Planck's constant $\hbar$ discretizing phase space. The physical
interpretation of this parameter is the smallest area in space that can be probed. The ordinary product between functions in non-commutative space is replaced by the $\star$-product, where the $\star$-product between
two arbitrary functions $f\left(x\right)$ and $g\left(x\right)$ is given by:
\begin{equation}\label{eq:2}
f\left(x\right)\star g\left(x\right)=f\left(x\right)\cdot g\left(x\right)+\left.\exp\left(\frac{i}{2}\,\theta^{ij}\,\partial_i\,\acute{\partial}_j\right)f\left(x\right) g\left(\acute{x}\right)\right\vert_{x=\acute{x}}\,.
\end{equation}

In this work we consider corrections to the gravitational force on the horizon of a Yukawa-Schwarzschild black hole due to thermodynamics properties in order to take into account non-commutativity for black holes. The black hole thermodynamics play an important role in the modern universe. The black hole thermodynamics provide a real connection between gravity and quantum mechanics. Recently there has been a lot of interest in the thermodynamics properties of black holes in non-commutative space-time \cite{4,5,6,7}. The thermodynamics properties of Yukawa-Schwarzchild black hole in commutative space was studied in Refs. \cite{8,9}. In this paper we apply some of these ideas to obtain thermodynamical properties for a Yukawa-Schwarzschild black hole in non-commutative space.

In the present study, we propose a deformed Yukawa-Schwarzschild black hole solution in non-commutative gauge theory of gravity. We apply the Bekenstein-Hawking method to compute the thermodynamical properties. The obtained qualitative results show that non-commutativity eliminates the divergent behavior of Hawking temperature and decreases the radius within which black holes can not radiate.

This paper is organized as follows. In section 2 we present the non-commutative gauge gravity metric by using Seiberg-Witten maps, following the approach of Ref. \cite{7}, and we calculate the corrected event horizon radius up to second order in the non-commutativity parameter. In section 3 we calculate the non-commutative temperature of the deformed Yukawa-Schwarzschild black hole, where we show that the non-commutative effects eliminate the divergent behavior of Hawking temperature and decrease the black hole radius to a new minimum limiting value. In section 4 we calculate the entropy of the deformed Yukawa-Schwarzschild black hole and show that the entropy modification of the black hole due to non-commutative space-time is negligible.

\section{Non-commutative gauge gravity metric for a Yukawa black hole}

In our previous work \cite{10}, we used the tetrad formalism and the inhomogeneous local Lorentz (Poincar\'{e}) transformations and imposed the invariance of the canonical non-commutative space-time commutation relations under the generalized local inhomogeneous Lorentz transformations by gauging $\mathrm{SO}(3,1)$, and by translating the enveloping algebra we constructed a theory of non-commutative general relativity. The generalized coordinate transformations are:
\begin{equation}\label{eq:3}
\hat{x}'^\mu=\hat{x}^\mu+\hat{\xi}^\mu(\hat{x})\,,
\end{equation}
which are compatible with the algebra given by \eqref{eq:1}. Under the change of coordinates \eqref{eq:3}, the non-commutative parameter $\hat{\xi}^\mu$ satisfies the following condition:
\begin{equation}
\hat{\xi}^{\mu}=\xi^{\mu}+\theta^{\mu\alpha}\partial_{\alpha }\xi^{\rho}\partial_{\rho}\,.
\end{equation}
Here we look for a mapping $\phi^{A}\rightarrow\hat{\phi}^{A}$ and $\lambda\rightarrow\hat{\lambda}$, where $\phi^{A}=(e_{\mu}^{a},\omega_{\mu}^{ab})$ is a generic field, $e_{\mu}^{a}$ and $\omega_{\mu}^{ab}$ are the vierbein and the gauge fields (spin connections) respectively (the Greek and Latin indices denote curved and tangent space-time respectively), and $\lambda=(\lambda_{P},\lambda_{G})$, where $\lambda _{P}$ and $\lambda _{G}$ are the local Poincar\'{e} and $\mathrm{U}(1)$ infinitesimal Lie-valued gauge-transformation parameters respectively, such that:
\begin{equation}\label{eq:trans}
\hat{\phi}^{A}\left(A\right)+\hat{\delta}_{\hat{\lambda}}\hat{\phi}^{A}\left(A\right)=\hat{\phi}^{A}\left( A+\delta _{\lambda }A\right),
\end{equation}
and
\begin{equation}
\lambda_{P}=\xi+\lambda_{L}\,,\qquad \hat{\lambda}_{P}=\hat{\xi}+\hat{\lambda}_{L}\,,
\end{equation}
with
\begin{equation}
\xi =\xi ^{\mu }\partial _{\mu }\,,\qquad \hat{\xi}=\hat{\xi}^{\mu}\partial _{\mu }\,,
\end{equation}
and
\begin{equation}
\lambda _{L}=\frac{1}{2}\lambda _{L}^{ab}S_{ab}\,,
\end{equation}
where $\xi $ and $\lambda _{L}$ are the local translation and generalized Lorentz infinitesimal Lie-valued gauge transformations respectively. The corresponding symmetry generators are denoted by $T_{a}$ (here we have a $\mathrm{U}(1)$ gauge group), $S_{ab}$ represent the generators of the Lorentz group and they satisfies the following relation:
\begin{equation}
\left[S_{ab},S_{cd}\right] =f_{abcd}^{ik}S_{ik}\,,
\end{equation}
with $f_{abcd}^{ik}$ being the structure constants of the Lorentz algebra.

In accordance with the general method of gauge theories, in the non-commutative space-time, we impose the infinitesimal local transformations given in eq \eqref{eq:trans}. Using these transformations, one can get at second order in the non-commutative parameter $\theta^{\mu \nu }$ the following Seiberg-Witten maps:
\begin{align}
\hat{e}_\mu^a=\,&e_\mu^a-\frac{i}{4}\theta^{\alpha\beta}\left(\omega_\alpha^{ac}\partial_\beta e_\mu^c+\left(\partial_\beta\omega_\mu^{ac}+R_{\beta\mu}^{ac}\right)e_\mu^c\right)\notag\\
&+\frac{1}{32}\theta ^{\alpha \beta }\theta ^{\gamma \delta }\left(2\left\{ R_{\delta \alpha }R_{\mu \beta }\right\} ^{ac}e_{\gamma}^{c}-\omega _{\gamma }^{ac}\left( D_{\beta }R_{\delta \mu }^{cd}+\partial
_{\beta }R_{\delta \mu }^{cd}\right) e_{\alpha }^{d}\right.  \notag \\
&\left. -\left\{ \omega _{\alpha }\left( D_{\beta }R_{\delta \mu }+\partial_{\beta }R_{\delta \mu }\right) \right\} ^{ad}e_{\gamma }^{d}-\partial_{\delta }\left\{ \omega _{\alpha }\left( \partial _{\beta }\omega _{\mu
}+R_{\beta \mu }\right) \right\} ^{ac}e_{\gamma }^{c}\right.  \notag \\
&\left. -\omega _{\gamma }^{ac}\partial _{\delta }\left\{ \omega _{\alpha}\partial _{\beta }e_{\mu }^{d}+\left( \partial _{\beta }\omega _{\mu}^{cd}+R_{\beta \mu }^{cd}\right) e_{\alpha }^{d}\right\} +\partial _{\alpha
}\omega _{\gamma }^{ac}\partial _{\beta }\partial _{\delta }e_{\mu}^{c}\right.  \notag \\
&\left. -\partial _{\beta }\left( \partial _{\delta }\omega _{\mu}^{ac}+R_{\delta \mu }^{ac}\right) \partial _{\alpha }e_{\gamma}^{c}-\left\{ \omega _{\alpha }\left( \partial _{\beta }\omega _{\gamma
}+R_{\beta \gamma }\right) \right\} ^{ac}\partial _{\delta }e_{\mu}^{c}\right.  \notag \\
&\left. -\partial_{\beta }\left( \partial _{\delta }\omega _{\mu }^{ac}+R_{\delta\mu }^{ac}\right) \left( \omega _{\alpha }^{cd}\partial _{\beta }e_{\gamma}^{d}+\left( \partial _{\beta }\omega _{\gamma }^{cd}+\left(R_{\left(\beta \gamma R_{\delta}^{cd}\right) e_{\alpha }^{d}}\right) \right) +\mathcal{O}\left( \theta^{3}\right)\right)\right.,\label{eq:SWM}
\end{align}
where
\begin{align}
R_{\beta \mu }^{ac}& =\partial _{\beta }\omega _{\mu }^{ac}-\partial _{\mu }\omega _{\beta }^{ac}+\left( \omega _{\mu }^{ab}\omega _{\beta }^{dc}-\omega _{\beta }^{ab}\omega _{\mu }^{dc}\right) \eta _{bd}\,, \\
R_{\beta \mu }& =R_{\beta \mu }^{ac}S_{ab}\,, \\
\omega _{\mu }& =\omega _{\mu }^{ab}S_{ab}\,, \\
\hat{\omega}_{\mu }& =\hat{\omega}_{\mu }^{ab}S_{ab}=\hat{\omega}_{k}^{ab}\ast \hat{e}_{\mu }^{k}S_{ab}\,, \\
\theta ^{\mu \nu }& =\hat{e}_{\ast a}^{\mu }\ast \hat{e}_{\ast b}^{\mu}\theta ^{ab}\,,
\end{align}
where the $\hat{e}_{\ast a}^{\mu }$ is the inverse-$\ast $ of the vierbein $\hat{e}_{\mu }^{a}$ defined as:
\begin{equation}
\hat{e}_{\mu }^{b}\ast \hat{e}_{\ast a}^{\mu }=\delta _{a}^{b}\,,
\end{equation}
and
\begin{equation}
\hat{e}_{\mu }^{a}\ast \hat{e}_{\ast a}^{\nu }=\delta _{\mu }^{\nu }\,.
\end{equation}

To begin, consider a non-commutative gauge theory with a charged scalar particle in the presence of an electrodynamic gauge field in a general curvilinear system of coordinates. We can write the action as:
\begin{equation}
\mathcal{S}=\frac{1}{2\kappa ^{2}}\int d^{4}x\,(\mathcal{L}_{G})\,,
\label{eq:action}
\end{equation}
where $\mathcal{L}_{G}$ is the pure gravity density defined as:
\begin{equation}
\mathcal{L}_{G}=\hat{e}\ast \hat{R}\,,
\end{equation}
with the deformed tetrad and scalar curvature given as:
\begin{align}
\hat{e}& =\det_{\ast}(\hat{e}_{\mu }^{a})\equiv \frac{1}{4!}\epsilon^{\mu \nu \rho \sigma }\varepsilon _{abcd}\hat{e}_{\mu }^{a}\ast \hat{e}_{\nu }^{b}\ast \hat{e}_{\rho }^{c}\ast \hat{e}_{\sigma }^{d}\,, \\
\hat{R}& =\hat{e}_{\ast a}^{\mu }\ast \hat{e}_{\ast b}^{\nu }\ast \hat{R}_{\mu \nu }^{ab}\,.
\end{align}
In the following, we consider a symmetric metric $\hat{g}_{\mu \nu }$, so that:
\begin{equation}\label{eq:22}
\hat{g}_{\mu \nu }=\frac{1}{2}(\hat{e}_{\mu }^{b}\ast \hat{e}_{\nu b}+\hat{e}_{\nu }^{b}\ast \hat{e}_{\mu b})\,.
\end{equation}
As a consequence, the first-order expansion in the non-commutative parameter $\theta ^{\alpha \beta }$ of the scalar curvature $\hat{R}$ and metric $\hat{g}_{\mu \nu }$ vanish. Thus $\hat{R}$ and $\hat{g}_{\mu \nu }$ can be
rewritten as:
\begin{align}
\hat{R}& =R+R^{(2)}+\mathcal{O}\left( \theta ^{3}\right), \\
\hat{g}_{\mu \nu }& =g_{\mu \nu }+g_{\mu \nu }^{(2)}+\mathcal{O}\left(\theta ^{3}\right),
\end{align}%
where $R^{(2)}=R_{rs}^{rs}$ was expressed explicitly in Ref. \cite{11}.

Using the Seiberg-Witten maps given in eq. \eqref{eq:SWM}, we can determine the deformed Yukawa Schwarzschild metric. To find this, first we have to obtain the corresponding components of the tetrad fields $e_{\mu }^{a}$ given by
Eqs. \eqref{eq:SWM}. With the definitions \eqref{eq:2} and \eqref{eq:22}, then it is possible to obtain the components of the deformed metric $\hat{g}_{\mu \nu }$. To simplify the calculations, we choose for the non-commutative antisymmetric metric $\theta^{\alpha \beta }$ the following:
\begin{equation}\label{eq:25}
\theta ^{\alpha \beta }=\left(\begin{array}{cccc}
0 & 0 & 0 & 0 \\
0 & 0 & \theta & 0 \\
0 & -\theta & 0 & 0 \\
0 & 0 & 0 & 0
\end{array}\right),\qquad \alpha\,,\beta =0,1,2,3\,,
\end{equation}
and for the dimensionless homogeneity in the non-commutative spaces, the coordinate operators must be redefined as: $x^\mu=\hat{x}^{\mu }/r_{0}$, where $\Theta=\theta/r_0^2$ ($r_0$ is the observed cosmic radius).

We follow the same steps outlined in Ref. \cite{12} and look for the non-commutative correction of the metric up to second order in $\Theta$. We choose the following diagonal tetrads:
\begin{align}
\underline{e}_{\mu }^{0}& =\left(\begin{array}{cccc}\left( 1-\frac{r_{s}}{r}\left( 1+\beta e^{-r/\lambda _{0}}\right) \right)^{1/2}, & 0, & 0, & 0\end{array}\right), \\
\underline{e}_{\mu }^{1}& =\left(\begin{array}{cccc}0, & \left( 1-\frac{r_{s}}{r}\left( 1+\beta e^{-r/\lambda _{0}}\right)\right) ^{-1/2}, & 0, & 0\end{array}\right), \\
\underline{e}_{\mu }^{2}& =\left(\begin{array}{cccc}0, & 0, & r, & 0\end{array}\right), \\
\underline{e}_{\mu }^{3}& =\left(\begin{array}{cccc}0, & 0, & 0, & r\sin \theta\end{array}\right).
\end{align}
The nonzero spin connections that follow from the null-torsion condition are:
\begin{align}
\omega _{\mu }^{12}& =\left( 0,-\left( 1-\frac{r_{s}}{r}\left( 1+\beta e^{-r/\lambda _{0}}\right) \right) ^{1/2},0,0\right), \\
\omega _{\mu }^{13}& =\left( 0,0,-\left( 1-\frac{r_{s}}{r}\left( 1+\beta e^{-r/\lambda _{0}}\right) \right) ^{1/2}\cos \theta ,0\right), \\
\omega _{\mu }^{23}& =\left( 0,0,-\cos \theta ,0\right), \\
\omega _{\mu }^{10}& =\left( 0,0,0,-\frac{r_{s}}{2r^{2}}\left( 1+\beta\left( 1+\frac{r}{\lambda _{0}}\right) e^{-r/\lambda _{0}}\right) \right),
\end{align}
where $r_{s}=2M/c^2$, with $M$ being the mass of the black hole, $\beta$ is the strength of the Yukawa correction and $\lambda_0$ represents the range of the Yukawa potential.

Using the Seiberg-Witten maps \eqref{eq:SWM}, and the choice \eqref{eq:25}, the non-zero components of the non-commutative tetrad fields $\hat{e}_{\mu }^{a}$ are:
\begin{align}
\hat{e}_{1}^{1}\,=&\frac{1}{\left( 1-\frac{\alpha }{r}\left( 1+\beta e^{-r/\lambda }\right) \right) ^{1/2}}-\frac{\alpha \left( 1+\beta \left( 1+\frac{r}{\lambda }+\frac{r^{2}}{2\tilde{\lambda}^{2}}\right) e^{-r/\lambda
}\right) }{8r^{3}\left( 1-\frac{\alpha }{r}\left( 1+\beta e^{-r/\lambda }\right) \right) }  \notag \\
&\left[ 1+\frac{\alpha \left( 1+\beta \left( 1+\frac{r}{\lambda }\right) e^{-r/\lambda }\right) ^{2}}{4r\left( 1-\frac{\alpha }{r}\left( 1+\beta e^{-r/\lambda }\right) \right) \left( 1+\beta \left( 1+\frac{r}{\lambda }+
\frac{r^{2}}{2\lambda ^{2}}\right) e^{-r/\lambda }\right) }\right] \Theta ^{2}+\mathcal{O}\left( \Theta ^{3}\right),\\
\hat{e}_{2}^{1} \,=& -\frac{i}{4}\left( 1-\frac{\alpha }{r}\left( 1+\beta e^{-r/\lambda }\right) \right) ^{1/2}\left[ 1+\frac{\alpha \left( 1+\beta \left( 1+\frac{r}{\lambda }\right) e^{-r/\lambda }\right) }{r\left( 1-\frac{
\alpha }{r}\left( 1+\beta e^{-r/\lambda }\right) \right) }\right] \Theta + \mathcal{O}\left( \Theta ^{3}\right),\\
\hat{e}_{2}^{2}\,=&r+\frac{1}{64r^{2}}\left[ 7\alpha \left( 1+\beta \left( 1+\frac{r}{\lambda }\right) e^{-r/\lambda }\right) \right. -  \notag \\
&\left. 24\alpha \left( 1+\beta \left( 1+\frac{r}{\lambda }+\frac{r^{2}}{2\lambda ^{2}}\right) e^{-r/\lambda }\right) \right] \Theta ^{2}+\mathcal{O}\left( \Theta ^{3}\right),\label{eq:36}\\
\hat{e}_{3}^{3}=\,&r\sin \theta -\frac{i}{4}\cos \theta \Theta +\frac{\alpha }{8r^{2}}\left[ \beta \left( 1+\frac{r}{\lambda }\right) e^{-r/\lambda}-\beta \left( 1+\frac{r}{\lambda }+\frac{r^{2}}{2\lambda ^{2}}\right)
e^{-r/\lambda }\right. + \notag\\
&\left. \frac{\left( 1+\beta \left( 1+\frac{r}{\lambda }\right) e^{-r/\lambda }\right) }{2\left( 1-\frac{\alpha }{r}\left( 1+\beta e^{-r/\lambda }\right) \right) }\left( \frac{\alpha \left( 1+\beta \left( 1+
\frac{r}{\lambda }\right) e^{-r/\lambda }\right) }{2r}-1\right) \sin \theta \right] \Theta ^{2}+\mathcal{O}\left( \Theta ^{3}\right),\\
\hat{e}_{0}^{0}\,=&\left( 1-\frac{\alpha }{r}\left( 1+\beta e^{-r/\lambda}\right) \right) ^{1/2}+\frac{\alpha }{8r^{3}}\left[ \frac{\alpha \left(1+\beta \left( 1+\frac{r}{\lambda }+\frac{r^{2}}{2\lambda ^{2}}\right)
e^{-r/\lambda }\right) }{r^{2}\left( 1-\frac{\alpha }{r}\left( 1+\beta e^{-r/\lambda }\right) \right) }\right. \times \notag \\
&\left( \frac{\alpha \left( 1+\beta \left( 1+\frac{r}{\lambda }\right) e^{-r/\lambda }\right) ^{2}}{4\left( 1-\frac{\alpha }{r}\left( 1+\beta e^{-r/\lambda }\right) \right) ^{1/2}}+3\left( 1+\beta \left( 1+\frac{r}{
\lambda }+\frac{r^{2}}{2\lambda ^{2}}\right) e^{-r/\lambda }\right) \right) \notag \\
&+\left. \frac{\alpha \left( 1+\beta \left( 1+\frac{r}{\lambda }+\frac{r^{2}}{2\lambda ^{2}}\right) e^{-r/\lambda }\right) ^{2}}{4r^{{}}\left( 1-\frac{\alpha }{r}\left( 1+\beta e^{-r/\lambda }\right) \right) ^{1/2}}\left( 1-
\frac{5\alpha \left( 1+\beta \left( 1+\frac{r}{\lambda }+\frac{r^{2}}{2\lambda ^{2}}\right) e^{-r/\lambda }\right) }{\left( 1-\frac{\alpha }{r}\left( 1+\beta e^{-r/\lambda }\right) \right) }\right) \right. \notag \\
&-\left( 1+\beta \left( 1+\frac{r}{\lambda }+\frac{r^{2}}{2\lambda ^{2}}\right) e^{-r/\lambda }\right) \left( 1-\frac{\alpha }{r}\left( 1+\beta e^{-r/\lambda }\right) \right) ^{1/2} \notag \\
&\left( 1- \frac{\alpha \left( 1+\beta \left( 1+\frac{r}{\lambda }\right)e^{-r/\lambda }\right) }{2r\left( 1-\frac{\alpha }{r}\left( 1+\beta e^{-r/\lambda }\right) \right) }\right) +\frac{3\alpha ^{2}\left( 1+\beta
\left( 1+\frac{r}{\lambda }+\frac{r^{2}}{2\lambda ^{2}}\right) e^{-r/\lambda }\right) ^{2}}{4r^{3}\left( 1-\frac{\alpha }{r}\left( 1+\beta e^{-r/\lambda }\right) \right) ^{3/2}} \notag\\
&\left.+ \frac{6}{r}\left( 1+\beta \left( 1+\frac{r}{\lambda }+\frac{r^{2}}{2\lambda ^{2}}+\frac{r^{3}}{6\lambda ^{3}}\right) e^{-r/\lambda }\right) \right] \Theta ^{2}+\mathcal{O}\left( \Theta ^{3}\right),
\end{align}
where $\alpha =r_s/r_0$ and $\lambda=\lambda _0/r_0$. Then, using the definition \eqref{eq:22}, we obtain the following non-zero components of the non-commutative metric $\hat{g}_{\mu \nu }$ up to second order of $\Theta$:
\begin{align}
\hat{g}_{11}&=\frac{1}{\left( 1-\frac{\alpha }{r}\left( 1+\beta e^{-r/\lambda }\right) \right) ^{1/2}}\left( 1-\frac{\alpha \beta e^{-r/\lambda }}{2\lambda ^{2}r}\Theta ^{2}\right) -  \notag \\
&\frac{\alpha \left( 1+\beta \left( 1+\frac{r}{\lambda }\right) e^{-r/\lambda }\right) }{r^{2}\left( 1-\frac{\alpha }{r}\left( 1+\beta e^{-r/\lambda }\right) \right) }\left[ 1+\frac{\alpha \left( 1+\beta \left(
1+\frac{r}{\lambda }\right) e^{-r/\lambda }\right) }{4r\left( 1-\frac{\alpha}{r}\left( 1+\beta e^{-r/\lambda }\right) \right) }\right] \Theta ^{2}+\mathcal{O}\left( \Theta ^{3}\right),\\
\hat{g}_{22} &=r^{2}\left( 1-\frac{3\alpha \beta e^{-r/\lambda }}{8\lambda^{2}r}\Theta ^{2}\right) +\frac{\left( 1-\frac{\alpha }{r}\left( 1+\beta e^{-r/\lambda }\right) \right) }{16}\left[ 1\right. -  \notag \\
&\left. \frac{13\alpha \left( 1+\beta \left( 1+\frac{r}{\lambda }\right) e^{-r/\lambda }\right) }{2r\left( 1-\frac{\alpha }{r}\left( 1+\beta e^{-r/\lambda }\right) \right) }+\frac{\alpha ^{2}\left( 1+\beta \left( 1+
\frac{r}{\lambda }\right) e^{-r/\lambda }\right) ^{2}}{r^{2}\left( 1-\frac{ \alpha }{r}\left( 1+\beta e^{-r/\lambda }\right) \right) ^{2}}\right] \Theta ^{2}+\mathcal{O}\left( \Theta ^{3}\right),\\
\hat{g}_{33} &=r^{2}\sin \theta ^{2}\left( 1-\frac{\alpha e^{-r/\lambda }}{8\lambda r}\Theta ^{2}\right) +\frac{\alpha \left( 1+\beta \left( 1+\frac{r}{\lambda }\right) e^{-r/\lambda }\right) }{2}\left[ 1-\frac{3\left( 1-\frac{2\alpha }{3r}\left( 1+\beta e^{-r/\lambda }\right) \right) }{4r\left( 1-\frac{\alpha }{r}\left( 1+\beta e^{-r/\lambda }\right) \right) }+\right. \notag \\
&\left. \frac{\alpha \left( 1+\beta \left( 1+\frac{r}{\lambda }\right)e^{-r/\lambda }\right) }{4r^{2}\left( 1-\frac{\alpha }{r}\left( 1+\beta e^{-r/\lambda }\right) \right) }\left( 1-\frac{1}{2\left( 1-\frac{\alpha }{r}
\left( 1+\beta e^{-r/\lambda }\right) \right) }\right) \sin \theta^{2}\right. +  \notag \\
&\left. \frac{\cos \theta ^{2}}{8\alpha \left( 1+\beta \left( 1+\frac{r}{\lambda }\right) e^{-r/\lambda }\right) }\right] \Theta ^{2}+\mathcal{O}\left( \Theta ^{3}\right),\label{eq:41}\\
\hat{g}_{00} &=-\left( 1-\frac{\alpha }{r}\left( 1+\beta e^{-r/\lambda}\right) \right) \left( 1+\frac{\alpha }{2r^{3}}\left( 1+\beta \left( 1+\frac{r}{\lambda }+\frac{r^{2}}{2\lambda ^{2}}\right) e^{-r/\lambda }\right)
\Theta ^{2}\right) +  \notag \\
&\frac{\alpha ^{2}\left( 1+\beta \left( 1+\frac{r}{\lambda }\right)e^{-r/\lambda }\right) }{8r^{4}}\left( 1+\beta \left( 1+\frac{r}{\lambda }+\frac{r^{2}}{2\lambda ^{2}}\right) e^{-r/\lambda }\right) \Theta ^{2}+
\mathcal{O}\left( \Theta ^{3}\right),\label{eq:42}
\end{align}

It is clear that, for $\Theta\rightarrow0$, we obtain the commutative Yukawa Schwarzschild solution. It is interesting to note that the deformed metric $\hat{g}_{\mu\nu}$ is spherically symmetry and is related to the choice of the non-commutative Yukawa parameter $\Theta$ as in \eqref{eq:25}. For such a black hole, we find that the event horizon in non-commutative space-time is where the non-commutative metric \eqref{eq:22} satisfies the following conditions:
\begin{equation}
\hat{g}_{00}=0\,.
\end{equation}
Let us analyze the condition for the existence of an event horizon in non-commutative space-time in the cases $\mathcal{O}(\lambda ^{-2})$ and $\mathcal{O}(1/r^4)$. At this approximation order, we can rewrite \eqref{eq:36} as follows:
\begin{equation}
\left( 1+\frac{\alpha \beta }{\lambda }\right) r^{3}-\alpha \left( 1+\beta\right) r^{2}+\frac{\alpha \left( 1+\beta \right) }{2}\left( 1+\frac{\alpha\beta }{\lambda }\right) \Theta ^{2}=0\,.
\end{equation}
We take $\delta =\left( 1+\alpha \beta/\lambda\right)$ and $\gamma =\alpha \left(1+\beta\right)$. Then this equation becomes:
\begin{equation}\label{eq:45}
\delta\,r^3-\gamma\,r^2+\frac{1}{2}\,\gamma\,\delta\,\Theta^2=0\,.
\end{equation}
The solution of this third-order polynomial equation is given in Ref. \cite{13}. Then the real root of the quadratic formula \eqref{eq:45} is the event horizon of the Yukawa Schwarzschild black hole in non-commutative space. The real root is given by:
\begin{align}
r_H^{NC}&=\frac{\gamma }{3\delta }\left[ 1+\sqrt[3]{1+\frac{\gamma }{4}\left( \frac{3\delta }{\gamma }\right) ^{3}\Theta ^{2}+i\sqrt{\frac{\gamma }{2}\left( \frac{3\delta }{\gamma }\right) ^{3}}\Theta }+\sqrt[3]{1+\frac{
\gamma }{4}\left( \frac{3\delta }{\gamma }\right) ^{3}\Theta ^{2}-i\sqrt{\frac{\gamma }{2}\left( \frac{3\delta }{\gamma }\right) ^{3}}\Theta }\right]\notag \\
&\simeq r_{H}^{C}\left( 1+3\left( 1+\frac{\alpha \beta }{\lambda }\right)\left( \frac{\Theta }{r_{H}^{C}}\right) ^{2}\right),
\end{align}
where $r_{H}^{C}=\alpha\left( 1+\beta \right)/\left(1+\alpha \beta/\lambda\right)$ is the event horizon in commutative case when $\Theta =0$. The effect of non-commutativity is exponentially small, which is reasonable to expect since at large distances space-time can be considered as a smooth classical manifold.

\section{Non-commutative temperature of Hawking radiation}

Let us now consider the black hole temperature. The corrected temperature of a black hole whose metric has been modified by a Yukawa potential term in non-commutative space-time is given by:
\begin{align}
\hat{T}_H&=\left(\frac{1}{4\pi }\left\vert \frac{dg_{00}}{dr}\right\vert \right) _{r=r_{H}^{NC}}\notag \\
&=T_{H}^{C}\left[ 1-\frac{15}{2}\left( 1+\frac{3\alpha \beta }{5\lambda }\right) \left( \frac{\Theta }{r_{H}^{C}}\right) ^{2}\right],
\end{align}
where $T_H^C=1/\left[4\pi\,r_H^C\left(1-\alpha\beta/\lambda\right)\right]$ is the semiclassical temperature of the Yukawa black hole in commutative spaces. For large black holes, i.e. when $\Theta/r_H^C\ll 1$, one recovers the standard result \cite{14,15,16} for the Hawking temperature:
\begin{equation}
T_H^C=\frac{1}{4\pi\,r_H^C\left(1-\frac{\alpha\beta}{\lambda}\right)}\simeq \frac{1}{4\pi\,r_H^C}\left(1+\frac{\alpha\beta}{\lambda}\right).
\end{equation}
\begin{figure}[ht]
\centering
\subfloat[$\beta_{\min}=1.38\times 10^{-11}$, $\lambda=6.081\times 10^6$]{\includegraphics[width=0.45\textwidth]{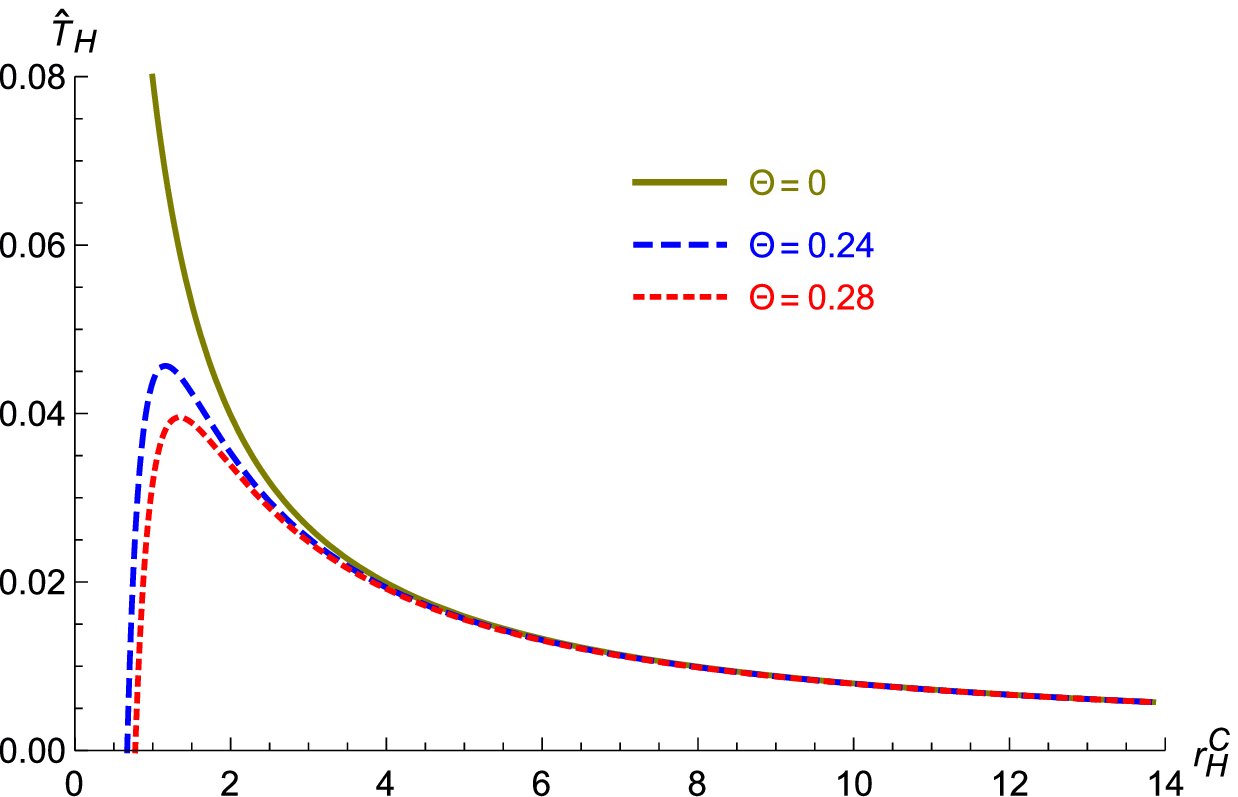}}\hfil
\subfloat[$\beta_{\min}=3.57\times 10^{-10}$, $\lambda=2.89 \times 10^{10}$]{\includegraphics[width=0.45\textwidth]{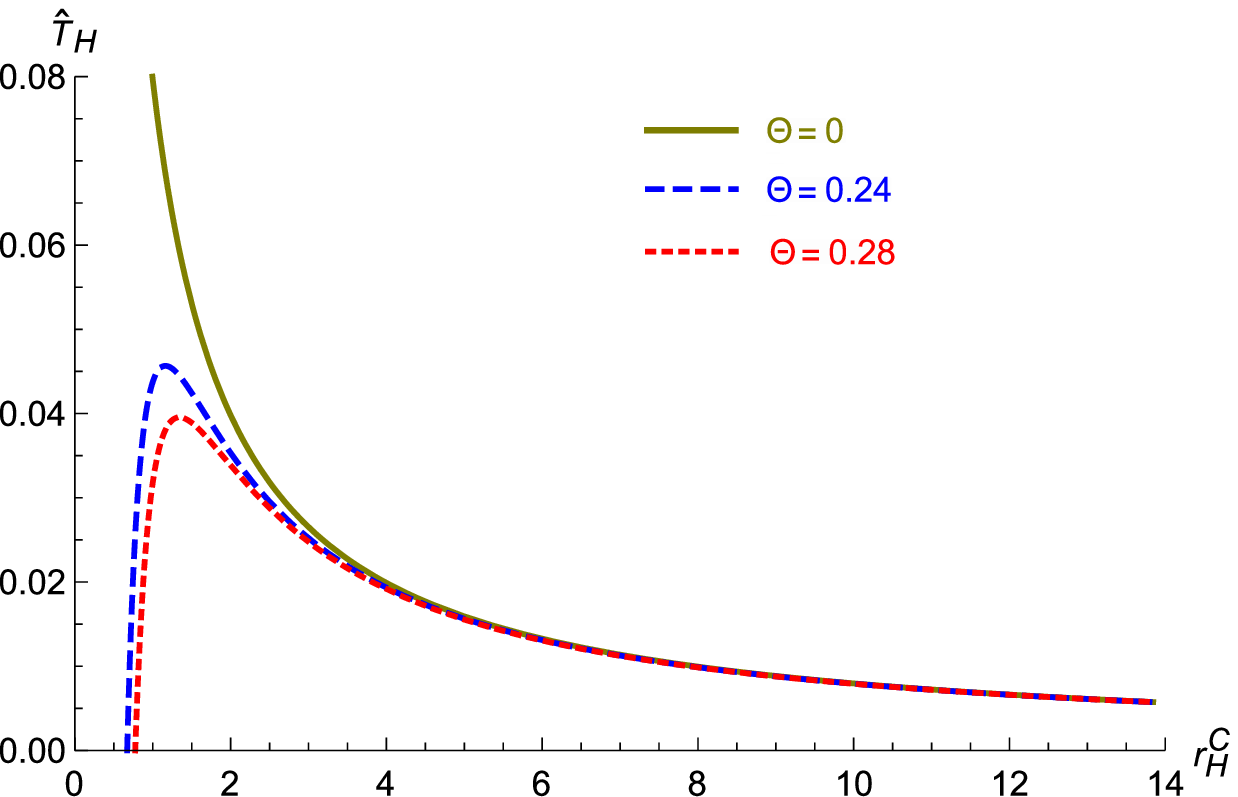}}
\caption{\label{fig:example1}Hawking non-commutative temperature $\hat{T}_{H}$ as a function of the horizon radius $r_{H}$.}
\end{figure}

We plot in figure \ref{fig:example1} the Hawking temperature for various non-commutative parameters. We see that the temperature (for dashed lines) reaches a maximum value $T_H^{\max}=2.28\times 10^{-2}/\Theta$ for $r_{H}=4.7\Theta$, and then decreases to zero as $r_{H}=2.7\Theta$. The commutative temperature (solid line) has a divergent behavior which is cured by non-commutativity.

The formula for $\hat{T}_{H}$ also leads to $\hat{T}_{H}=0$ for
\begin{equation}
r_{H}^{C}=r^{O}=\sqrt{\frac{15}{2}\left(1+\frac{3\alpha \beta }{5\lambda }\right) }\Theta \simeq 2.7\Theta\,.
\end{equation}
The non-commutative temperature $\hat{T}_{H}$ reaches its maximum for
\begin{equation}
r_{H}^{C}=3\sqrt{\frac{5}{2}\left( 1+\frac{3\alpha \beta }{5\lambda }\right) }\Theta \simeq 4.7\Theta\,.
\end{equation}
These results differ from what is found in the standard (commutative space) case where $T_{H}^{C} $ diverges as $r_{H}^{C}$ goes to zero.

On the thermodynamics side, in the case of non-commutative space, we see that non-commutativity eliminates the divergent behavior of Hawking temperature. As a result, there is a new maximum temperature that a black hole can reach before turning to zero for a new small non-zero radius $r^{O}$. In the region $r_{H}^{C}<r^{O}$ there is a black hole with negative $\hat{T}_{H}$, and this has no physical sense. Thus, in commutative space, the effect of vacuum fluctuations increases the value of Hawking temperature. On the contrary, the value of the Hawking temperature increases to maximum value and then decreases to zero, so that there are no processes of particle-antiparticle creation on the external surface of the black hole with a radius $r_{H}^{C}=r^{O}$. The obtained results show that non-commutativity decreases the radius within which black holes can not radiate. These results correspond to what the authors of Ref. \cite{17} expected.

\section{Entropy correction on the non-commutative horizon of the black hole}

The black hole entropy depends on the black hole event horizon $A_{H}$ with initial formula:
\begin{equation}
S=\frac{A_{H}}{4}\,.
\end{equation}
The surface area of the Yukawa black hole in non-commutative space-time at second order in $\Theta$ is as follows:
\begin{align}
A_{H} &=4\pi \left( r_{H}^{NC}\right) ^{2}  \notag \\
&=4\pi \left( \frac{\alpha \left( 1+\beta \right) }{\left( 1+\frac{\alpha \beta }{\lambda }\right) }\right) ^{2}\left( 1+6\left( 1+\frac{\alpha \beta}{\lambda }\right) \left( \frac{\Theta }{r_{H}^{C}}\right) ^{2}\right).
\end{align}
With the help of Eqs. \eqref{eq:41} and \eqref{eq:42}, the corrected non-commutative entropy becomes:
\begin{align}
S_{H}^{NC}=&\pi \left( \frac{\alpha \left( 1+\beta \right) }{\left( 1+\frac{\alpha \beta }{\lambda }\right) }\right) ^{2}+6\pi \left( 1+\frac{\alpha \beta }{\lambda }\right) \Theta ^{2} \notag \\
=&S_{H}^{C}+6\pi \left( 1+\frac{\alpha \beta }{\lambda }\right) \Theta ^{2}\,,
\end{align}
where
\begin{equation}
S_{H}^{C}=\pi \left( \frac{\alpha \left( 1+\beta \right) }{\left( 1+\frac{\alpha \beta }{\lambda }\right) }\right) ^{2}\simeq \pi \alpha^{2}\left( 1+\beta \right) ^{2}\left( 1-2\frac{\alpha \beta }{\lambda }
\right).
\end{equation}
\begin{figure}[ht]
\centering
\subfloat[$\beta_{\min}=1.38\times 10^{-11}$, $\lambda=6.081\times 10^{6}$]{\includegraphics[width=0.45\textwidth]{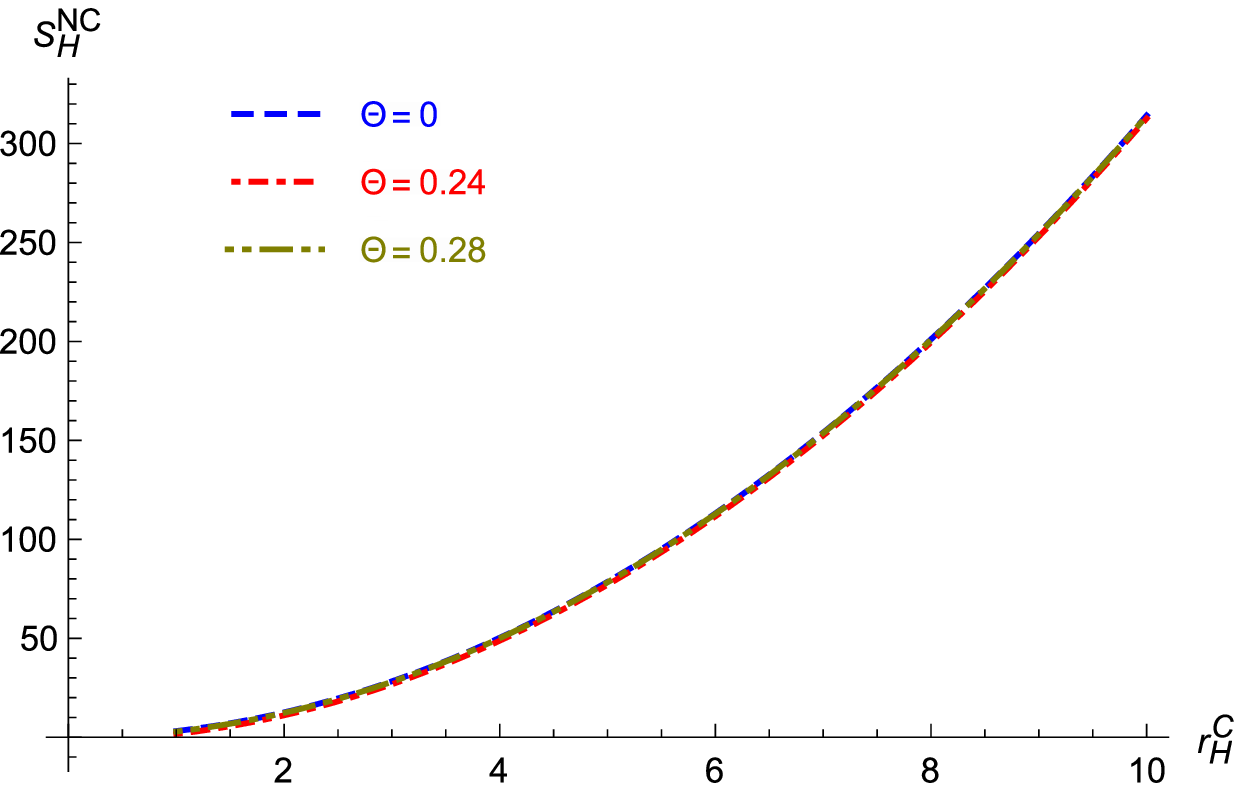}}\hfil
\subfloat[$\beta_{\min}=3.57\times 10^{-10}$, $\lambda=2.89\times 10^{10}$]{\includegraphics[width=7cm]{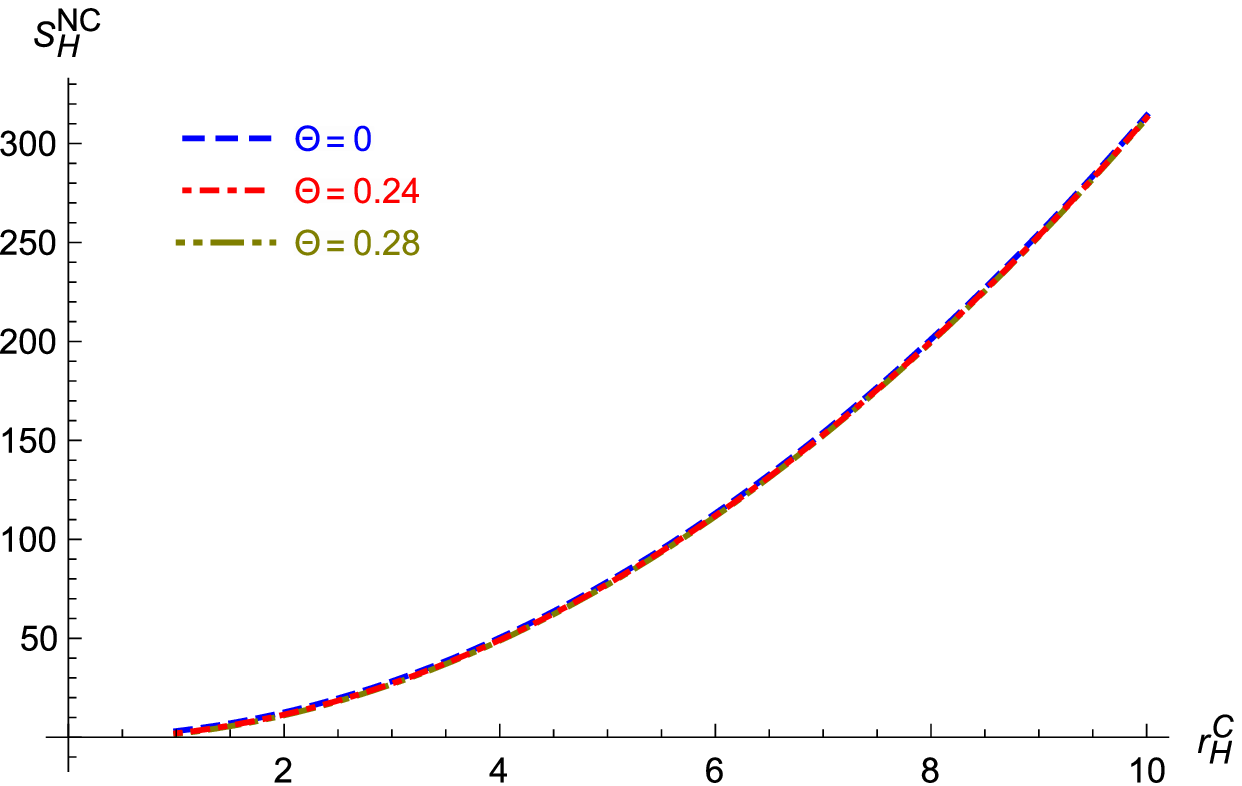}}
\caption{\label{fig:example}Entropy correction on the non-commutative horizon of the black hole $\hat{S}_{H}^{NC}$ as a function of the horizon radius $r_{H}^{C}$.}
\end{figure}
See figrure \ref{fig:example}. A similar analysis can be performed over non-commutativity, finding that for small values of $\Theta $, variations over $S_{H}^{NC}$ and $S_{H}^{C}$ are negligible. This finding is consistent with other results on corrections to the black hole entropy \cite{18,19}. This confirms our confidence in quantitative mechanical corrections to Newton's potential and their conclusions.

\section{Conclusion}

We analyzed thermodynamics properties of non-commutative Yukawa-Schwarzschild black hole. We obtained the corrected non-commutative temperature and entropy of these black hole solutions. We noted that the non-commutativity modifies the thermodynamics properties of the Yukawa-Schwarzschild black hole. First, we found the non-commutative temperature and that these non-commutative corrections eliminate the divergent behavior of Hawking temperature and they lead to a new maximum temperature that the black hole can reach before turning to zero at a new minimum value for the black hole radius. The values of these results were predicted in Ref \cite{15}.

Then we obtained the corrections for the entropy of the Yukawa-Schwarzschild black hole due to the small non-commutative parameter. These corrections are negligible. When the non-commutative parameter $\Theta\rightarrow 0$ these corrections vanish, and hence we have derived the ordinary results for Yukawa Schwarzschild black hole. The obtained results support the idea of introducing a correction on the Newton's potential for quantum gravity.

\end{document}